\documentclass[
  aps,
  pra,
  preprint,
  amsmath,
  amssymb,
  superscriptaddress,
  longbibliography
]{revtex4-2}

\usepackage[T1]{fontenc}
\usepackage{graphicx}
\usepackage{bm}
\usepackage{placeins}
\usepackage{hyperref}
\usepackage{orcidlink}
\hypersetup{
  colorlinks=true,
  linkcolor=black,
  citecolor=black,
  urlcolor=black,
  pdftitle={Spectral structure and controlled energy linearization of Feshbach effective Hamiltonians in a molecular quantum quench},
  pdfauthor={Georgii V. D'yakonov},
  pdfkeywords={Feshbach effective Hamiltonian, matrix-valued spectral measure, sudden quantum quench, self-energy, full configuration interaction}
}

\newcommand{\Eh}{E_{\mathrm h}}
\newcommand{\HT}{\mathrm{HT}}
\newcommand{\HeH}{{}^{3}\mathrm{HeH}^{+}}
\newcommand{\Psii}{\Psi_{\mathrm i}}
\newcommand{\Ran}{\operatorname{Ran}}
\newcommand{\supp}{\operatorname{supp}}

\newcommand{\norm}[1]{\left\lVert #1\right\rVert}

\begin{document}

\title{Spectral structure and controlled energy linearization of Feshbach effective Hamiltonians
in a molecular quantum quench}

\author{Georgii\ V.\ D'yakonov \orcidlink{0009-0009-7048-2862}}
\affiliation{Faculty of Chemistry, Lomonosov Moscow State University,
Moscow 119991, Russia}
\date{August 19, 2026}

\begin{abstract}
We establish spectral conditions under which a first-order energy linearization of a
projected Feshbach Hamiltonian is controlled.  When the support of the induced
self-energy spectral measure lies entirely above or below the target energy window,
the self-energy curvature and linearization remainder have fixed sign in the
semidefinite (L{\"o}wner) order.  The remainder then decreases monotonically under
model-space enlargement by complete complementary-space spectral subspaces, and the
minimax linearization energy is determined by the endpoint norms.  Far-separated
asymptotics and the near-boundary divergence caused by an approaching pole are
obtained explicitly.  The theory is validated by a numerically exact sudden-quench
calculation for the two-electron system
\(\mathrm{HT}\rightarrow{}^3\mathrm{HeH}^+\): all 56 nontrivially coupled
basis--geometry cases have support separated above the target window, while seven
cc-pVTZ cases are exactly decoupled by construction and serve as numerical control
cases.  The same remainder controls the projected resolvent and, through source
dressing, the full prepared-state response, with resolvent amplification governing
the near-pole sensitivity.  These results provide a practical criterion for
controlled downfolding in molecular electronic structure and for final-state
analyses in neutrino-mass experiments.
\end{abstract}

\maketitle

\section{Introduction}
\label{sec:introduction}

Projection of a Hamiltonian onto a finite model space produces an exact reduced
operator whose dependence on the spectral parameter carries the influence of the
eliminated sector.  For orthogonal projectors $P$ and $Q=I-P$, the Feshbach effective
Hamiltonian is
\begin{equation}
 \begin{aligned}
 H_{\mathrm{eff}}(z)&=H_{PP}+\Sigma_P(z),\\
 \Sigma_P(z)&=H_{PQ}(z-H_{QQ})^{-1}H_{QP}.
 \end{aligned}
 \label{eq:intro-heff}
\end{equation}
Here $H_{PP}=PHP$, $H_{PQ}=PHQ$, $H_{QP}=QHP$, and $H_{QQ}=QHQ$.
Feshbach and L{\"o}wdin partitioning establish the exact energy-dependent reduction
generated by a Hilbert-space projection
\cite{Feshbach1958,Lowdin1951Partitioning,Lowdin1962Projection}.
Projection-based effective Hamiltonians have long been used in atomic and molecular
structure, including multiply excited resonances and molecular model Hamiltonians
\cite{Bylicki1989Feshbach,SoliverezGaglianoArteca1985}.
Bloch--Horowitz and energy-independent model-space constructions, together with
unitary block diagonalization, provide related effective-interaction frameworks
\cite{BlochHorowitz1958,KuoLeeRatcliff1971,Bravyi2011SchriefferWolff}.
Modern Feshbach--Schur formulations place the same projection map within rigorous
perturbation theory for self-adjoint operators \cite{DussonSigalStamm2021}.
Multistate perturbative and downfolding methods formulate related reduced-space
eigenvalue problems in molecular electronic structure
\cite{Finley1998MSCASPT2,Angeli2001NEVPT,Tenno2013EffectiveHamiltonian,
Bauman2019Downfolding,Lang2020HermitianQDNEVPT,HouChen2023}.
Energy-dependent and nonlocal molecular effective descriptions also arise in
continuum-coupled vibronic dynamics \cite{DvorakHoufekCizek2022}.
Controlled effective-Hamiltonian representations also remain an active problem in
quantum dynamics, including robust Hamiltonian engineering and perturbative
extraction of coherent reduced dynamics
\cite{ChenCory2026,CollaBreuerGasbarri2025}.

A first-order energy linearization replaces the exact dependence of
$H_{\mathrm{eff}}(E)$ by a first-order expansion about a linearization energy
$E_0$.  Its accuracy over a target energy window $I=[E_-,E_+]$ is governed by the
self-energy linearization remainder.  The spectral theorem identifies the relevant
eliminated-space object: the positive matrix-valued measure induced on
$\mathcal H_P=\Ran P$.  Its support and residue matrices determine the sign,
magnitude, and energy variation of the remainder.  A nearby complementary-space
eigenvalue contributes according to its residue; a subspace with zero residue is
absent from the self-energy pole expansion.

In one-particle Green's-function theory, dynamical self-energies have likewise been
recast through finite or static effective Hamiltonians, with pole and intruder-state
structure entering the resulting quasiparticle problem
\cite{MoninoLoos2022,ScottBackhouseBooth2023,MarieLoos2023}.
Throughout this article, $\Sigma_P$ denotes the Feshbach self-energy generated by the
$P/Q$ Hilbert-space projection.

Sudden nuclear transmutation supplies a physical preparation of final electronic
states.  In the sudden approximation the electronic state is unchanged at the
instant of the event, and direct projection onto final eigenstates gives the
transition probabilities \cite{Migdal1941,LandauLifshitzQM,WilliamsKoonin1983}.
Final-state distributions in tritium-bearing molecules form part of molecular
beta-decay theory
\cite{Kolos1985,Fackler1985,Claxton1992,SaenzFroelich1997a,
SaenzFroelich1997,Jonsell1999,Saenz2000,Doss2006}.
The bound-ion branching ratio of the HT isotopologue has been measured directly
\cite{LinEtAl2020HT}.  Molecular final-state distributions remain central to
contemporary neutrino-mass analyses \cite{Bodine2015,Schneidewind2024}.  The
two-electron process
$\HT\rightarrow\HeH$ permits full configuration interaction in the total-spin
singlet sector and a complete finite-basis decomposition of $H_{QQ}$, while the
basis-set representation and internuclear distance remain controllable parameters.

Support of the self-energy spectral measure entirely on one side of the target
window gives the self-energy linearization remainder a definite sign.  Enlargement
by complete complementary-space spectral subspaces gives a L{\"o}wner-monotone
reduction of that remainder.  The same one-sided support reduces the uniform minimax
problem to endpoint norms, while the exactly solvable single-pole model gives the
far-separated and near-boundary asymptotics.  All 56
nontrivially coupled $\HT\rightarrow\HeH$ finite-basis realizations satisfy these
sufficient conditions through nonzero-residue support separated above the target
window.  The exact projected-resolvent identity propagates the same remainder to
poles associated with the target states and to prepared-state spectral weights.

\begin{figure*}[t]
 \includegraphics[width=\textwidth]{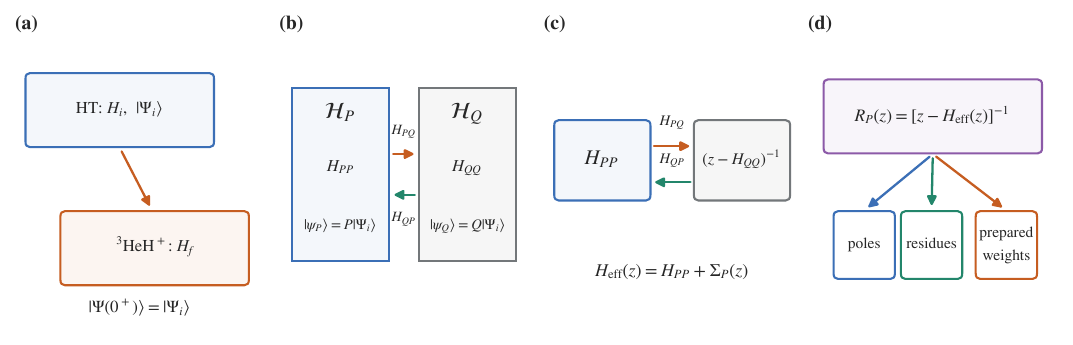}
 \caption{\label{fig:quench-feshbach}
 From sudden preparation to the exact Feshbach effective Hamiltonian.
 (a) The electronic state is unchanged at the instant of the
 $\HT\rightarrow\HeH$ quench.  (b) The final Hilbert space is decomposed into
 $\mathcal H_P$ and $\mathcal H_Q$.  (c) Elimination of $Q$ yields the
 energy-dependent self-energy
 $\Sigma_P(z)=H_{PQ}(z-H_{QQ})^{-1}H_{QP}$.
 (d) The projected resolvent contains the poles, residues, and prepared-state
 spectral weights.}
\end{figure*}

\section{Spectral structure and controlled energy linearization of Feshbach effective Hamiltonians}
\label{sec:theory}

\subsection{Exact projected resolvent and spectral measure}

Let $P^2=P$, $Q=I-P$, $P+Q=I$, and $PQ=0$.  The full Hilbert space is
$\mathcal H=\mathcal H_P\oplus\mathcal H_Q$, with
$\mathcal H_P=\Ran P$ and $\mathcal H_Q=\Ran Q$, and the Hamiltonian has block form
\begin{equation}
 H=
 \begin{pmatrix}
 H_{PP} & H_{PQ}\\
 H_{QP} & H_{QQ}
 \end{pmatrix}.
 \label{eq:block-hamiltonian}
\end{equation}
For the resolvent $R(z)=(z-H)^{-1}$, the Schur complement gives
\begin{equation}
 \boxed{
 R_P(z)=PR(z)P=
 [zI_P-H_{PP}-\Sigma_P(z)]^{-1}}
 \label{eq:projected-resolvent}
\end{equation}
whenever $z-H_{QQ}$ is invertible.  Equation~\eqref{eq:projected-resolvent} is an
exact identity on $\mathcal H_P$.

The spectral theorem resolves the complementary block as
\begin{equation}
 H_{QQ}=\int_{\mathbb R}\epsilon\,dE_Q(\epsilon),
 \qquad
 dM(\epsilon)=H_{PQ}\,dE_Q(\epsilon)\,H_{QP}.
 \label{eq:induced-measure}
\end{equation}
Matrix-valued spectral measures and their Cauchy-transform representations provide a
standard framework for spectral problems involving self-adjoint operators and
finite-rank perturbations \cite{GesztesyTsekanovskii2000,LiawTreil2020}.
The induced measure $dM(\epsilon)$ is positive semidefinite and acts on
$\mathcal H_P$.  With the resolvent convention used here, the Feshbach self-energy is
the matrix-valued Cauchy transform
\begin{equation}
 \boxed{
 \Sigma_P(z)=\int_{\mathbb R}
 \frac{dM(\epsilon)}{z-\epsilon}.}
 \label{eq:spectral-self-energy}
\end{equation}
In a finite orbital space,
\begin{equation}
 \begin{aligned}
 H_{QQ}&=\sum_\alpha\epsilon_\alpha\Pi_\alpha,\\
 W_\alpha&=H_{PQ}\Pi_\alpha H_{QP}\succeq0,\\
 \Sigma_P(z)&=\sum_\alpha\frac{W_\alpha}{z-\epsilon_\alpha}.
 \end{aligned}
 \label{eq:finite-spectral-self-energy}
\end{equation}
The complementary-space eigenvalues $\epsilon_\alpha$ locate possible poles and the
residue matrices $W_\alpha$ determine their model-space action, including induced
off-diagonal couplings.  A zero-residue spectral subspace, $W_\alpha=0$, contributes no term to
Eq.~\eqref{eq:finite-spectral-self-energy}.  The relevant spectral set is therefore
$\supp M$.

\subsection{First-order energy-linearized representation and one-sided spectral support}

For $E_0\in I=[E_-,E_+]$, define the first-order energy-linearized self-energy
\begin{equation}
 \Sigma_P^{[1]}(E;E_0)=
 \Sigma_P(E_0)+(E-E_0)\Sigma_P'(E_0).
 \label{eq:local-self-energy}
\end{equation}
Substitution in the projected eigenvalue equation gives
\begin{equation}
 H_0^{\mathrm{loc}}c=E N_0c,
 \label{eq:local-generalized-eigenproblem}
\end{equation}
with
\begin{equation}
 \begin{aligned}
 H_0^{\mathrm{loc}}
 &=H_{PP}+\Sigma_P(E_0)-E_0\Sigma_P'(E_0),\\
 N_0&=I_P-\Sigma_P'(E_0).
 \end{aligned}
 \label{eq:local-operators}
\end{equation}
For real energies outside $\supp M$,
\begin{equation}
 \Sigma_P'(E)=-\int_{\mathbb R}
 \frac{dM(\epsilon)}{(E-\epsilon)^2}\preceq0,
 \label{eq:sigma-prime}
\end{equation}
so the norm operator satisfies $N_0\succ0$.  If $|\Phi_n\rangle$ is a normalized
exact eigenstate and $c_n=P|\Phi_n\rangle$, reconstruction of its complementary
component yields the Feshbach normalization
\begin{equation}
 1=\langle c_n|[I_P-\Sigma_P'(E_n)]|c_n\rangle.
 \label{eq:feshbach-normalization}
\end{equation}

The self-energy linearization remainder is
\begin{equation}
 \Delta\Sigma(E;E_0)=
 \Sigma_P(E)-\Sigma_P(E_0)-(E-E_0)\Sigma_P'(E_0).
 \label{eq:remainder-definition}
\end{equation}
Taylor's integral formula gives
\begin{equation}
 \Delta\Sigma(E;E_0)
 = (E-E_0)^2\int_0^1(1-t)
 \Sigma_P''[E_0+t(E-E_0)]\,dt.
 \label{eq:integral-remainder}
\end{equation}
Combining this identity with Eq.~\eqref{eq:spectral-self-energy} yields
\begin{equation}
 \boxed{\begin{aligned}
 \Delta\Sigma(E;E_0)
 &=\int_{\mathbb R}dM(\epsilon)
 \frac{(E-E_0)^2}
 {(E_0-\epsilon)^2(E-\epsilon)}.
 \end{aligned}}
 \label{eq:spectral-remainder}
\end{equation}
In finite dimensions, the integral in Eq.~\eqref{eq:spectral-remainder} becomes a
sum over $W_\alpha$.  Its scalar kernel
\begin{equation}
 K(E,E_0;\epsilon)=
 \frac{(E-E_0)^2}{(E_0-\epsilon)^2(E-\epsilon)}
 \label{eq:kernel}
\end{equation}
contains the complete energy dependence of each positive residue matrix.

The second energy derivative is
\begin{equation}
 \Sigma_P''(E)=2\int_{\mathbb R}
 \frac{dM(\epsilon)}{(E-\epsilon)^3}.
 \label{eq:self-energy-curvature}
\end{equation}
For self-energy spectral support separated above the target window
(hereafter upper-separated support),
\begin{equation}
 \supp M\subset(E_+,\infty),
 \qquad
 \delta_+=\inf\supp M-E_+>0,
 \label{eq:upper-separation}
\end{equation}
the denominator in Eqs.~\eqref{eq:spectral-remainder} and
\eqref{eq:self-energy-curvature} is negative throughout $I$.  Hence
\begin{equation}
 \boxed{
 \Sigma_P''(E)\preceq0,
 \qquad
 \Delta\Sigma(E;E_0)\preceq0
 \quad (E,E_0\in I).}
 \label{eq:upper-sign}
\end{equation}
For lower-separated support,
$\supp M\subset(-\infty,E_-)$ and
$\delta_-=E_--\sup\supp M>0$, the corresponding relations are
$\Sigma_P''(E)\succeq0$ and $\Delta\Sigma(E;E_0)\succeq0$.
The order symbol $A\preceq B$ denotes the L{\"o}wner order: $B-A$ is positive
semidefinite.

\begin{figure*}[t]
 \includegraphics[width=\textwidth]{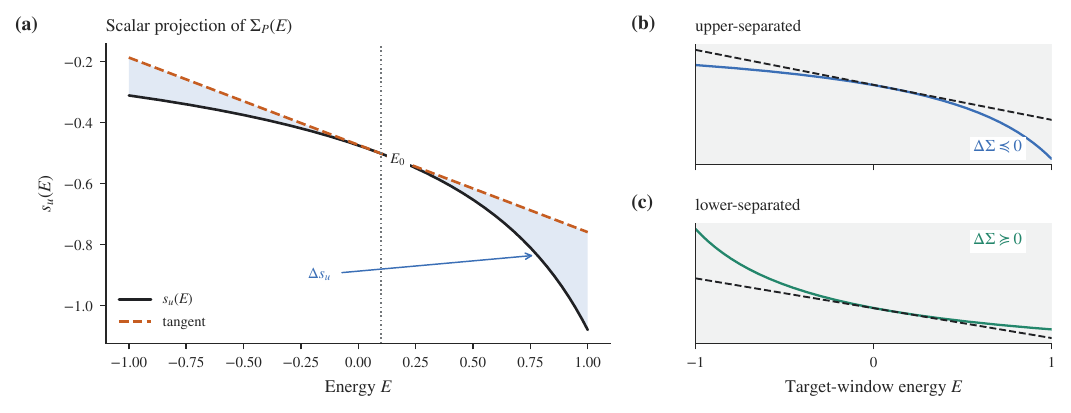}
 \caption{\label{fig:self-energy-geometry}
Curvature of the self-energy.
(a) Scalar projection $s_u(E)$ and its tangent at $E_0$ for the explicit positive
 spectral model; the shaded separation is
 $\Delta s_u=\langle u|\Delta\Sigma|u\rangle$.
 (b,c) Upper and lower spectral separation give negative- and positive-semidefinite
 remainders, respectively.}
\end{figure*}

Figure~\ref{fig:self-energy-geometry} visualizes a normalized scalar projection
$s_u(E)=\langle u|\Sigma_P(E)|u\rangle$ of the matrix-valued self-energy.  The plotted
curves are generated from an explicit positive spectral model.  The matrix statement
is Eq.~\eqref{eq:upper-sign}.

The sign relation also controls spectral-projector enlargement of the model space.
Let $P_1$ be the initial projector, $Q_1=I-P_1$, and let
$\Pi_S\preceq Q_1$ be a complete spectral projector of $H_{Q_1Q_1}$.  Define
\begin{equation}
 \begin{aligned}
  P_2&=P_1+\Pi_S, & Q_2&=Q_1-\Pi_S,\\
  \mathcal H_{P_2}&=\mathcal H_{P_1}\oplus\Ran\Pi_S.
 \end{aligned}
 \label{eq:spectral-enlargement-space}
\end{equation}
Spectral invariance gives
$\Pi_S H_{Q_1Q_1}Q_2=0$; the transferred sector therefore has no self-energy coupling
to $Q_2$.  On the common space $\mathcal H_{P_2}$, embed the old remainder as
\begin{equation}
 \widetilde{\Delta\Sigma}_{P_1}(E;E_0)
 =\Delta\Sigma_{P_1}(E;E_0)\oplus0_{\Pi_S}.
 \label{eq:embedded-remainder}
\end{equation}
For upper-separated support, removal of the contribution associated with the positive
residue on $\Pi_S$ gives
\begin{equation}
 \widetilde{\Delta\Sigma}_{P_1}(E;E_0)
 \preceq \Delta\Sigma_{P_2}(E;E_0)\preceq0,
 \label{eq:loewner-enlargement}
\end{equation}
and therefore
\begin{equation}
 \begin{aligned}
  \norm{\Delta\Sigma_{P_2}(E;E_0)}_2
  &\leq \norm{\widetilde{\Delta\Sigma}_{P_1}(E;E_0)}_2\\
  &=\norm{\Delta\Sigma_{P_1}(E;E_0)}_2.
 \end{aligned}
 \label{eq:norm-enlargement}
\end{equation}
The ordering applies to enlargement by complete spectral subspaces of the
complementary Hamiltonian.

\subsection{Minimax linearization and asymptotic limits}

Define the uniform operator-norm linearization remainder and its minimax energy by
\begin{equation}
 \Phi(E_0)=\sup_{E\in I}\norm{\Delta\Sigma(E;E_0)}_2,
 \qquad
 E_0^\star\in\underset{E_0\in I}{\arg\min}\,\Phi(E_0).
 \label{eq:minimax-definition}
\end{equation}
For upper-separated support, the positive operator
$-\Delta\Sigma(E;E_0)$ grows in L{\"o}wner order as $E$ moves from $E_0$ toward either
endpoint.  Consequently,
\begin{equation}
 \boxed{
 \Phi(E_0)=\max\left\{
 \norm{\Delta\Sigma(E_-;E_0)}_2,
 \norm{\Delta\Sigma(E_+;E_0)}_2
 \right\}.}
 \label{eq:endpoint-minimax}
\end{equation}
Write
\begin{equation}
 \begin{aligned}
  L(E_0)&=\norm{\Delta\Sigma(E_-;E_0)}_2,\\
  U(E_0)&=\norm{\Delta\Sigma(E_+;E_0)}_2.
 \end{aligned}
 \label{eq:endpoint-branches}
\end{equation}
For $\epsilon>E_+$ and $E_0\in I$, the endpoint kernel coefficients satisfy
\begin{equation}
 \begin{aligned}
  \frac{d}{dE_0}\frac{(E_0-E_-)^2}{(\epsilon-E_0)^2(\epsilon-E_-)}
  &=\frac{2(E_0-E_-)}{(\epsilon-E_0)^3}\geq0,\\
  \frac{d}{dE_0}\frac{(E_+-E_0)^2}{(\epsilon-E_0)^2(\epsilon-E_+)}
  &=-\frac{2(E_+-E_0)}{(\epsilon-E_0)^3}\leq0.
 \end{aligned}
 \label{eq:endpoint-kernel-monotonicity}
\end{equation}
Integration against $dM(\epsilon)\succeq0$ therefore makes $L(E_0)$ nondecreasing
and $U(E_0)$ nonincreasing on $I$.
Every intersection $L(E_0)=U(E_0)$ is a minimax point.
We use
\begin{equation}
 \xi^\star=\frac{E_0^\star-E_-}{\Omega},
 \qquad \Omega=E_+-E_-,
 \label{eq:normalized-minimax}
\end{equation}
where $\Omega$ is the target-window width and is distinct from the residue matrices
$W_\alpha$.
If $H_{PQ}=0$, then $\Sigma_P(z)=0$, $\Delta\Sigma(E;E_0)=0$, and
$\Phi(E_0)=0$ for every $E_0\in I$.  Every point of $I$ is then a minimizer, and no
unique coordinate $\xi^\star$ is assigned to an exactly decoupled control case.

If $\supp M\subset[E_++\delta,\infty)$, Eq.~\eqref{eq:spectral-remainder} gives the
uniform bound
\begin{equation}
 \norm{\Delta\Sigma(E;E_0)}_2
 \leq \frac{(E-E_0)^2}{\delta^3}
 \norm{H_{PQ}H_{QP}}_2,
 \label{eq:separation-bound}
\end{equation}
and hence
\begin{equation}
 \Phi(E_0)\leq
 \frac{\Omega^2}{\delta^3}\norm{H_{PQ}H_{QP}}_2.
 \label{eq:window-bound}
\end{equation}
The coefficient follows directly from the exact kernel; choosing the midpoint
replaces $\Omega^2$ by $\Omega^2/4$ in this bound.

For poles above the target window, set $h=E-E_0$ and
$\Delta_\alpha=\epsilon_\alpha-E_0>0$.  In the far-separated regime
\(\Omega\ll\delta\), we use the formal asymptotic expansion of the exact kernel
\begin{equation}
 \begin{aligned}
 \Delta\Sigma(E;E_0)
 &=-h^2\mathcal M_3-h^3\mathcal M_4
 -h^4\mathcal M_5-\cdots,\\
 \mathcal M_n(E_0)&=\sum_\alpha
 \frac{W_\alpha}{\Delta_\alpha^n}.
 \end{aligned}
 \label{eq:moment-expansion}
\end{equation}
In this notation $\Sigma_P(E_0)=-\mathcal M_1$ and
$\Sigma_P'(E_0)=-\mathcal M_2$.  The far-separated regime
$\Omega\ll\delta$ therefore gives
$\Delta\Sigma=O(\Omega^2/\delta^3)$ at fixed residue scale.
Under the formal weak-coupling scaling $H_{PQ}\mapsto\lambda V_{PQ}$ with fixed
leading-order $H_{QQ}$ spectrum,
\begin{equation}
 \begin{aligned}
 W_\alpha&=O(\lambda^2),&
 \Sigma_P&=O(\lambda^2/\delta),\\
 N_0-I_P&=O(\lambda^2/\delta^2),&
 \Delta\Sigma&=O(\lambda^2\Omega^2/\delta^3).
 \end{aligned}
 \label{eq:weak-coupling}
\end{equation}

The single-pole case is exactly solvable and has a unique minimizer.  Let
$\epsilon=E_++\delta$ and let $W_1\succeq0$ be its residue matrix.  The minimax
coordinate and remainder are
\begin{equation}
 \boxed{\begin{aligned}
 \xi_{\mathrm{sp}}^\star
 &=\frac{\sqrt{\Omega+\delta}}
 {\sqrt{\Omega+\delta}+\sqrt{\delta}}
 \end{aligned}}
 \label{eq:single-pole-xi}
\end{equation}
\begin{equation}
 \boxed{\begin{aligned}
 \Phi_{\mathrm{sp}}^\star
 &=\norm{W_1}_2
 \left[
 \frac{1}{\sqrt{\delta}}-
 \frac{1}{\sqrt{\Omega+\delta}}
 \right]^2
 \end{aligned}}
 \label{eq:single-pole-phi}
\end{equation}
For $\delta\gg\Omega$,
\begin{align}
 \xi_{\mathrm{sp}}^\star
 &=\frac12+\frac{\Omega}{8\delta}
 -\frac{\Omega^2}{16\delta^2}
 +\frac{5\Omega^3}{128\delta^3}+\cdots,
 \label{eq:far-xi}\\
 \Phi_{\mathrm{sp}}^\star
 &=\norm{W_1}_2\left[
 \frac{\Omega^2}{4\delta^3}
 -\frac{3\Omega^3}{8\delta^4}
 +\frac{29\Omega^4}{64\delta^5}+\cdots\right].
 \label{eq:far-phi}
\end{align}
Thus $E_0^\star=E_{\mathrm{mid}}+\Omega^2/(8\delta)
-\Omega^3/(16\delta^2)+\cdots$, with
$E_{\mathrm{mid}}=(E_-+E_+)/2$.  In the near-boundary limit
$\delta\rightarrow0^+$,
\begin{align}
 E_+-E_0^\star
 &=\sqrt{\Omega\delta}-\delta
 +\frac{\delta^{3/2}}{2\sqrt{\Omega}}+\cdots,
 \label{eq:near-position}\\
 \Phi_{\mathrm{sp}}^\star
 &=\norm{W_1}_2\left[
 \frac{1}{\delta}-\frac{2}{\sqrt{\Omega\delta}}
 +\frac{1}{\Omega}+\cdots\right].
 \label{eq:near-phi}
\end{align}
The approach $E_+-E_0^\star\sim\sqrt{\Omega\delta}$ places the linearization energy
near the approaching pole, while
$\Phi_{\mathrm{sp}}^\star\sim\norm{W_1}_2/\delta$ expresses the loss of uniform
first-order accuracy.

\begin{figure*}[t]
 \includegraphics[width=\textwidth]{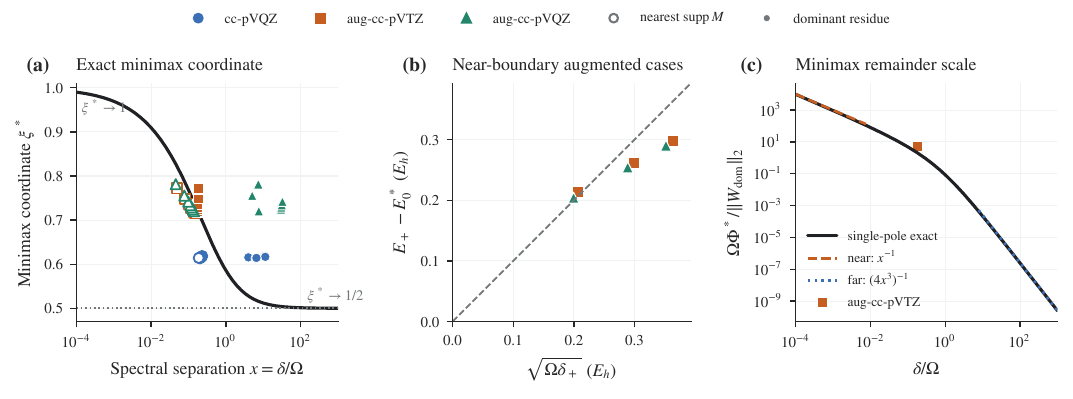}
 \caption{\label{fig:minimax-asymptotics}
 Minimax structure and asymptotic limits.
 (a) Exact single-pole coordinate $\xi_{\mathrm{sp}}^\star$ versus
 $\delta/\Omega$, together with molecular minimax points.  Open and filled symbols
 use the nearest point of $\supp M$ and the largest-residue spectral point,
 respectively, to define the horizontal coordinate.
 (b) Near-boundary augmented-basis cases compared with
 $E_+-E_0^\star=\sqrt{\Omega\delta_+}$.
 (c) Exact dimensionless single-pole remainder and asymptotic forms, drawn over domains
 where their relative error is below 20\%; aug-cc-pVTZ points display the multipole
 molecular comparison.}
\end{figure*}

\section{Molecular realization in a sudden quantum quench}
\label{sec:molecular}

\subsection{Physical preparation and finite-basis realization}

The nuclear event changes the fixed-nuclei electronic Hamiltonian from $H_i$ for
$\HT$ to $H_f$ for $\HeH$.  In the sudden approximation,
\begin{equation}
 |\Psi(0^+)\rangle=|\Psii\rangle,
 \qquad
 H_f|\Phi_n^{(f)}\rangle=E_n|\Phi_n^{(f)}\rangle,
 \label{eq:sudden-state}
\end{equation}
and the final-state probabilities are
\begin{equation}
 p_n=|\langle\Phi_n^{(f)}|\Psii\rangle|^2.
 \label{eq:sudden-probability}
\end{equation}
The three lowest prepared cc-pVTZ daughter singlets at the reference distance
$R_0=0.7563593751$~\AA\ define $P^{(3)}$.  Their total unrenormalized probability is
$0.825551$ in the species-adapted parent--daughter construction.  Cross-basis
determinant overlaps transport this subspace to the other basis-set representations;
its singular values and prepared-state projection weights are reported in the
Supplemental Material~\cite{SupplementalMaterial}.

The fixed-nuclei Hamiltonian contains two electrons on two centers and is diagonalized
in the total-spin singlet sector by full configuration interaction, giving exact
diagonalization of each finite-basis Hamiltonian \cite{Olsen1988,Sun2020}.  The scan
contains cc-pVXZ and aug-cc-pVXZ ($X={\rm D,T,Q}$) basis-set representations
\cite{Dunning1989,KendallDunningHarrison1992} and the independent def2-SVP,
def2-TZVPP, and def2-QZVPP family \cite{WeigendAhlrichs2005}.  Species-adapted
parent--daughter overlaps use the corresponding cross-basis AO overlap matrix;
fixed-AO control calculations retain the same H-labelled Gaussian functions during
the charge change.  Seven
geometries cover $0.8\leq R/R_0\leq2.0$.  Complete diagonalization of $H_{QQ}$ gives
all $\epsilon_\alpha$ and $W_\alpha$ for every finite-basis calculation.
Implementation, mapping, tolerances, and complete scans are given in the
Supplemental Material~\cite{SupplementalMaterial}.

\subsection{Molecular self-energy spectral structure}

The nine basis-set representations and seven geometries produce 63 cases.  All 56
nontrivially coupled cases satisfy
$\supp M\subset(E_+,\infty)$ over the three-state target window.  The seven cc-pVTZ
cases have $H_{PQ}=0$ and serve as exactly decoupled control cases.  Several augmented
representations contain eigenvalues of $H_{QQ}$ inside $I$ whose residue matrices
vanish.  These levels leave $\Sigma_P(z)$ analytic across the target window.

\begin{figure*}[t]
 \includegraphics[width=\textwidth]{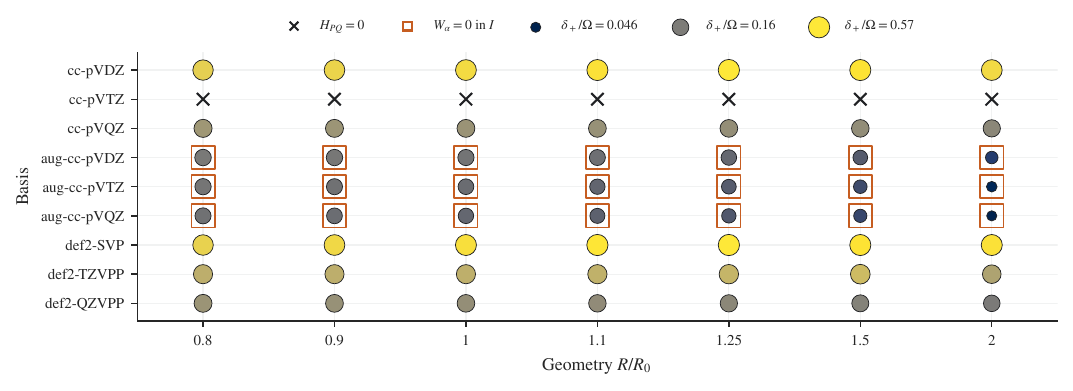}
 \caption{\label{fig:molecular-support}
 The basis--geometry map of $\delta_+/\Omega$ for nontrivially coupled cases.
 Crosses identify $H_{PQ}=0$ control cases and open squares mark cases with a zero-residue
 level inside $I$.}
\end{figure*}

At $R=R_0$, the two representative cases give
\begin{equation*}
 \begin{gathered}
 \mathrm{cc\text{-}pVQZ}:\\[-3pt]
 I/\Eh=[-2.977287,-1.709780],\\[-1pt]
 (\delta_+/\Eh,E_0^\star/\Eh,\xi^\star)
   =(0.278327,-2.195626,0.616692);\\[3pt]
 \mathrm{aug\text{-}cc\text{-}pVTZ}:\\[-3pt]
 I/\Eh=[-2.975162,-1.738375],\\[-1pt]
 (\delta_+/\Eh,E_0^\star/\Eh,\xi^\star)
   =(0.161134,-2.080610,0.723287).
 \end{gathered}
\end{equation*}
The corresponding uniform remainders are $2.058\times10^{-3}$ and
$2.128\times10^{-2}$~$\Eh$.  Figure~\ref{fig:molecular-support} resolves the
nonzero-residue spectral subspaces and their transfer across the entire scan.

The nearest point of $\supp M$ and the largest-residue spectral point, defined by
$\alpha_{\mathrm{dom}}\in\arg\max_\alpha\norm{W_\alpha}_2$, generally define
different spectral scales.  In cc-pVQZ at $R_0$, the nearest normalized distance is
$\delta_+/\Omega=0.219586$, whereas the largest-residue point lies much farther from
the window.  In aug-cc-pVTZ the corresponding coordinates are $0.130284$ and
$0.172669$, and the single-pole curve tracks $\xi^\star$ more closely when the
largest-residue coordinate is used.  The molecular self-energy remains a multipole
operator, as displayed by deviations from the single-pole curve in
Fig.~\ref{fig:minimax-asymptotics}(a,c).

Stretching the augmented-basis cases moves the nearest nonzero-residue support toward
the target boundary.  At $R/R_0=2.0$, the ratios
$(E_+-E_0^\star)/\sqrt{\Omega\delta_+}$ are 1.032 for aug-cc-pVTZ and 1.023 for
aug-cc-pVQZ, consistent with the near-boundary scale in
Eq.~\eqref{eq:near-position}.  The calculation varies basis and geometry; the formal
coupling parameter in Eq.~\eqref{eq:weak-coupling} is held fixed.

Model-space enlargement from $P^{(3)}$ to a five-dimensional space $P^{(5)}$
transfers two complete complementary-space spectral subspaces.  The five-dimensional
spaces were fixed by a pre-existing selection rule given in the Supplemental
Material~\cite{SupplementalMaterial} and
are retained unchanged here.  At $R_0$, the sampled minimax remainder decreases from
$2.06\times10^{-3}$ to $3.15\times10^{-4}$~$\Eh$ in cc-pVQZ and from
$2.15\times10^{-2}$ to $3.45\times10^{-3}$~$\Eh$ in aug-cc-pVTZ.  These cases realize
the L{\"o}wner-monotone spectral-subspace enlargement of
Eq.~\eqref{eq:loewner-enlargement}.

\section{Projected spectral reconstruction}
\label{sec:reconstruction}

\begin{figure*}[!t]
 \includegraphics[width=\textwidth]{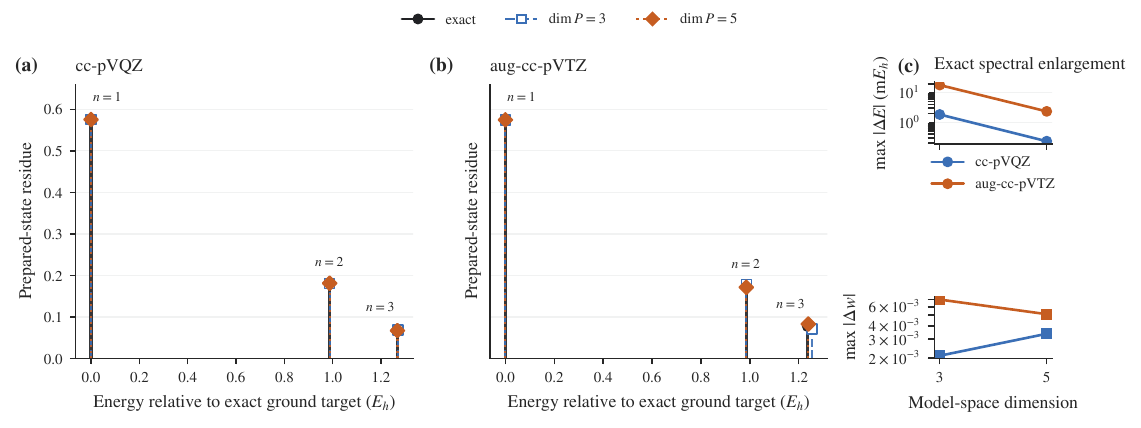}
 \caption{\label{fig:poles-residues}
 Projected poles and prepared-state spectral weights at $R_0$.
 (a,b) Exact full-source target sticks $w_n^{\mathrm{full}}$ and linearized
 projected-source sticks $\widetilde w_n^P$ for energy-linearized effective theories
 with $\dim P=3$ and 5 in cc-pVQZ and aug-cc-pVTZ.  Energies are measured from the
 exact target-state ground energy in each basis-set representation; stick heights are
 unrenormalized prepared-state spectral weights.
 (c) Maximum pole errors and
 $\max_n|\Delta w_n|=\max_n|\widetilde w_n^P-w_n^{\mathrm{full}}|$ are shown on
 separate axes under spectral-subspace enlargement.}
\end{figure*}

The energy-linearized projected resolvent associated with
Eqs.~\eqref{eq:local-generalized-eigenproblem} and \eqref{eq:local-operators} is
\begin{equation}
 R_P^{[1]}(z)=[zN_0-H_0^{\mathrm{loc}}]^{-1}.
 \label{eq:local-resolvent}
\end{equation}
Resolvent subtraction gives the exact identity
\begin{equation}
 \boxed{
 R_P(z)-R_P^{[1]}(z)=
 R_P(z)\Delta\Sigma(z;E_0)R_P^{[1]}(z).}
 \label{eq:resolvent-remainder-identity}
\end{equation}
The remainder identity also controls the full prepared-state response once the source
is dressed.  Decomposing the prepared state as
\(|\psi_P\rangle=P|\Psii\rangle\), \(|\psi_Q\rangle=Q|\Psii\rangle\), and defining
the dressed sources
\begin{subequations}
\begin{align}
|\eta_R(z)\rangle &= |\psi_P\rangle + H_{PQ}(z-H_{QQ})^{-1}|\psi_Q\rangle,\\
\langle\eta_L(z)| &= \langle\psi_P| + \langle\psi_Q|(z-H_{QQ})^{-1}H_{QP},
\end{align}
\end{subequations}
the exact full response is
\(g(z)=\langle\Psii|R(z)|\Psii\rangle
 =\langle\psi_Q|(z-H_{QQ})^{-1}|\psi_Q\rangle
 +\langle\eta_L(z)|R_P(z)|\eta_R(z)\rangle\).
Replacing the exact projected resolvent \(R_P\) by its energy-linearized approximation
\(R_P^{[1]}\)
and using Eq.~\eqref{eq:resolvent-remainder-identity} gives
\begin{equation}
\boxed{
g(z)-g_{\mathrm{dress}}^{[1]}(z)
=
\langle\eta_L(z)|R_P(z)\,\Delta\Sigma(z;E_0)\,R_P^{[1]}(z)|\eta_R(z)\rangle,
}
\end{equation}
where
\(g_{\mathrm{dress}}^{[1]}(z)=
\langle\psi_Q|(z-H_{QQ})^{-1}|\psi_Q\rangle
+\langle\eta_L(z)|R_P^{[1]}(z)|\eta_R(z)\rangle\).
Thus the same remainder \(\Delta\Sigma\) that governs the projected resolvent also
controls the full prepared-state response, with resolvent and dressed-source factors
setting the final sensitivity scale.  Near a pole, their norms amplify a small
uniform remainder.
Numerical validation of the source-dressing reconstruction is reported in the
Supplemental Material~\cite{SupplementalMaterial}.

For the energy-linearized generalized eigenvalue problem, the eigenvectors satisfy
\begin{equation}
 H_0^{\mathrm{loc}}u_j=\widetilde E_j N_0u_j,
 \qquad
 u_i^\dagger N_0u_j=\delta_{ij}.
 \label{eq:generalized-normalization}
\end{equation}
This norm-operator normalization gives the linearized projected-source residue
$\widetilde w_n^P=|u_n^\dagger\psi_P|^2$.  At the corresponding exact target pole,
define the residue of the full prepared-state response (the full-source residue) and
the exact projected-source residue as
$w_n^{\mathrm{full}}=|\langle\Phi_n|\Psii\rangle|^2$ and
$w_n^P=|\langle\Phi_n|P|\Psii\rangle|^2$, respectively.  Using the
Feshbach-normalized projected eigenvector in Eq.~\eqref{eq:feshbach-normalization}
gives the latter expression.  Figure~\ref{fig:poles-residues} plots
$w_n^{\mathrm{full}}$ for the exact sticks and
$\widetilde w_n^P$ for the linearized projected-source sticks, without renormalizing
$P|\Psii\rangle$.  The signed residue difference is
\begin{equation}
 \begin{aligned}
 \Delta w_n&=\widetilde w_n^P-w_n^{\mathrm{full}}\\
 &=\underbrace{(\widetilde w_n^P-w_n^P)}_{\text{Hamiltonian linearization}}
 +\underbrace{(w_n^P-w_n^{\mathrm{full}})}_{\text{source projection}}.
 \end{aligned}
 \label{eq:residue-error-decomposition}
\end{equation}

The total residue differences $\max_n|\Delta w_n|$ are decomposed into the separately
recorded Hamiltonian-linearization and source-projection terms in
Eq.~\eqref{eq:residue-error-decomposition}.  In cc-pVQZ this maximum changes from
$2.102\times10^{-3}$ to
$3.363\times10^{-3}$ while the maximum target-pole error decreases from
$1.840\times10^{-3}$ to $2.30\times10^{-4}$~$\Eh$.  In aug-cc-pVTZ the maximum
total residue difference changes from $6.996\times10^{-3}$ to
$5.115\times10^{-3}$ while the
maximum target-pole error decreases from $1.8150\times10^{-2}$ to
$2.343\times10^{-3}$~$\Eh$.

Direct inversion and the exact Feshbach projected resolvent agree within
$1.1\times10^{-11}$ relative error at the fixed complex-energy validation points.
The source-dressing identity is satisfied at the same accuracy.  The matrix and
scalar forms of Eq.~\eqref{eq:resolvent-remainder-identity} agree within
$8.9\times10^{-10}$ relative error, and isolated contour residues reproduce
algebraic residues to
$1.2\times10^{-16}$.  Broadened response curves preserve the exact algebraic
residues while displaying the resolvent amplification near poles.

\section{Conclusions}
\label{sec:conclusions}

The Feshbach self-energy is the matrix-valued Cauchy transform, in the resolvent
convention $z-\epsilon$, of a positive spectral measure induced on the model space.
Its support and residue matrices govern the energy dependence of the exact projected
Hamiltonian.  One-sided support fixes the sign of the self-energy curvature and
linearization remainder in the semidefinite (L{\"o}wner) order.  Enlargement by
complete complementary-space spectral subspaces gives a L{\"o}wner-monotone reduction
of the remainder and reduces the uniform minimax problem to an endpoint balance.  The
exactly solvable single-pole limit connects the far-separated
$\Omega^2/\delta^3$ behavior to the near-boundary $1/\delta$ divergence.

Across $0.8\leq R/R_0\leq2.0$, all 56 nontrivially coupled
$\HT\rightarrow\HeH$ finite-basis calculations have nonzero-residue self-energy
support separated above the target window; the seven cc-pVTZ cases are exactly
decoupled by construction.  The relevant energy-separation scale is set by
$\supp M$.  Zero-residue complementary-space levels inside the target window leave
$\Sigma_P(z)$ analytic there.  Five-dimensional spectral-subspace enlargement
reduces the uniform remainder and the maximum target-pole errors in both principal
basis-set representations, directly verifying the L{\"o}wner-monotone improvement.

The projected-resolvent identity
$R_P-R_P^{[1]}=R_P\Delta\Sigma R_P^{[1]}$ shows that the same remainder controls the
projected spectral response.  Through source dressing, it also governs the full
prepared-state response, with near-pole resolvent factors setting the final
amplification scale.  The spectral conditions established here thus provide a
practical criterion for controlled energy-linearized downfolding in molecular
electronic-structure calculations and for final-state analyses in neutrino-mass
experiments.

\FloatBarrier
\begin{widetext}
\section*{Declarations}

\noindent\textbf{Conflict of interest.}
The author declares no conflicts of interest.

\noindent\textbf{Data availability.}
The matrix-valued spectral data, numerical validation tables, and underlying 
computational results supporting the findings of this study are openly available 
in Zenodo~\cite{dyakonov_2026_zenodo}.
\end{widetext}

\bibliography{references_final}

\end{document}